\documentclass{iaus}
\def\CH2DP{CH$_2$D$^+$}
\def\chd{CH$_2$D$^+$}
\def\ch{CH$_3^+$}
\def\hd{H$_2$D$^+$}
\def\h3{H$_3^+$}
\def\etal{{\it et~al\/}.}
\usepackage{graphicx}

\title[Interstellar CH$_2$D$^+$] 
{A Search for Interstellar CH$_2$D$^+$}

\author[Alwyn Wootten \& Barry E. Turner]   
{Alwyn Wootten$^1$ \and Barry E. Turner$^1$}

\affiliation{$^1$National Radio Astronomy Observatory \\ 520 Edgemont Road,
Charlottesville, Virginia 22903, USA \\ email: {\tt awootten@nrao.edu} }

\pubyear{2008}
\volume{251}  
\pagerange{1-4}
\jname{Organic Matter in Space}
\editors{A.C. Editor, B.D. Editor \& C.E. Editor, eds.}
\begin{document}

\maketitle

\begin{abstract}
We report on a search for Interstellar \CH2DP. Four transitions occur in easily accessible portions of the spectrum; we report on emission at the frequencies of these transitions toward high column density star-forming regions.  While the observations can be interpreted as being consistent with a detection of the molecule, further observations will be needed to secure that identification.  The \chd rotational spectrum has not been measured to high accuracy.   Lines are weak, as the dipole moment induced by the inclusion of deuterium in the molecule is small.  Astronomical detection is favored by observations toward strongly deuterium-fractionated sources.  However, enhanced deuteration is expected to be most significant at low temperatures.  The sparseness of the available spectrum and the low excitation in regions of high fractionation make secure identification of CH$_2$D$^+$ difficult.  Nonetheless, owing to the importance of CH$_3^+$ to interstellar chemistry, and the lack of rotational transitions of that molecule owing to its planar symmetric structure, a measure of its abundance would provide key data to astrochemical models.
\keywords{astrochemistry, line: identification, ISM: molecules, radio lines: ISM, submillimeter}
\end{abstract}

\firstsection 
\section{Introduction}

The symmetric species \ch is a reactant of extreme importance in interstellar organic
chemistry, as it initiates the formation of more complex hydrocarbons.  Unfortunately \ch cannot be observed through its rotational lines, as it is symmetric.  \chd is asymmetric, however, and emits a rotational spectrum.  Because the deuterium is bound to \chd more tightly than hydrogen to
\ch, and because the binding energies are similar to typical molecular cloud
temperatures, \chd becomes more abundant relative to its undeuterated
counterpart in cold clouds.  This binding energy is higher for the deuterium in \chd than for that in \hd, so that in warmer clouds, \chd remains heavily
fractionated to much higher temperatures than \hd.  This expectation is
borne out by observations of its deuterated derivatives.  Since DCO$^+$ derives from reaction of CO with
\hd, the temperature dependence of its fractionation mimics that of [\hd]/[\h3],
as demonstrated in observations of a cross-section of clouds by \cite[Wootten, Loren and Snell (1982)]{WoottenLorSn82}.The observed [DCO$^+$]/[HCO$^+$] ratio reaches high values only in the coldest clouds; DCO$^+$ is practically unobservable
in a warm cloud such as OMC1.    Persistence of a high [DCN]/[HCN] ratio to high temperatures, as observed by 
\cite[Greason (1986)]{Greason86} and discussed by \cite[Wootten (1987)]{Wootten87}
 by the same effect mimics the behavior of [\chd]/[\ch].   Both
because it remains abundant at relatively high temperatures, and
because rotational lines are accessible, \chd is an ideal candidate for
observation and confirmation of deuterium fractionation theory.
The three accessible mm-wave transitions of \chd are J$_{K_{-1}K_{1}}$=1$_{01}$-0$_{00}$ 
near 280 GHz, 2$_{11}$-2$_{12}$ near 200 GHz, and 1$_{10}$-1$_{11}$ near 67 GHz.
All should be detectable in warm molecular clouds, given model abundances
of \ch, the expected degree of deuteration, and the temperatures of
appropriate sources.


Good frequency estimates for rotational lines are available from \cite[Rosslein \etal 1991]{Ross91} and \cite[Jagod \etal 1992]{Jagod92}.   The estimated accuracy of the frequencies is estimated by  \cite[Rosslein \etal 1991]{Ross91} to be  $\pm 2 \times 10^{-5}$ (1 $\sigma$) times the frequency.  Four lines are relatively easily observed by radiotelescopes; we have attempted to detect all four.  The lowest frequency but highest excitation line, at 23.01595 GHz, was sought in 1992 April at the NRAO 43m telescope; no emission was detected.   The J$_{K_{-1}K_{1}}=1_{01}\rightarrow0_{00}$ line  was observed at a center frequency of 278.69162 GHz in 1992 at the 10.4m CSO telescope on Mauna Kea, Hawaii.   The J$_{K_{-1}K_{1}}=1_{01}\rightarrow0_{00},  1_{11}\rightarrow1_{10}$ and $ 2_{11}\rightarrow2_{12}$
lines  were observed at center frequencies of 278.69162 GHz, 67.27371 GHz and 201.76264 GHz 
at various times between 1992 April and 1997 September at the 12m NRAO telescope at Kitt Peak, Arizona.  
At the frequencies where emission from \chd was expected, emission was detected in Ori HC, NGC6334N, SgrB2OH N and W51M.  The detections were most convincing for NGC 6334N.  
The NGC6334N region is complex, with several sources appearing within the ~30Ó beam of our telescopes. The dominant sources within our beam include SMA1, SMA2, SMA3 and SMA4 (Hunter 2006).  In high excitation ammonia studies, SMA1 and SMA2 dominate the emission; emission was not detected toward SMA3 and SMA4. 
Column densities on the order of N$_{Tot}=2 \times 10^{14}$ are indicated for T$_{rot}\sim$50K.  The study of deuterium fractionation in warm clumps recently published by Roueff, Parise and Herbst (2007) predicts [\chd]/[\ch] $\sim$0.03 for 50K, declining by a factor of a few as T rises.  Our data would suggest then that N$_{Tot}$(\ch) exceeds $7 \times 10^{15}$.  Additional laboratory data is urgently needed to secure the identity of the lines we report.
 \begin{figure}[b]
\begin{center}
 \includegraphics[width=2.3in]{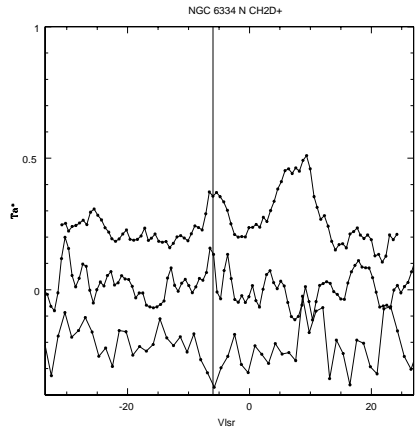} 
 \caption{Spectrum of NGC6334N in the vicinity of the J$_{K_{-1}K_{1}}=2_{11}\rightarrow2_{12}$ (201 GHz, top, 12m SSB data, center, offset by 0.2 K),  $(1_{01}\rightarrow0_{00}$ (278 GHz, center, CSO DSB data, upper) and J$_{K_{-1}K_{1}}=1_{11}\rightarrow1_{10}$ (67GHz, lower, 12m SSB data, lower, offset by -0.2K) lines of \chd.  The center spectrum has been multiplied by a factor of four.  The pointing center was 17$^h$ 17$^m$ 32.0$^s$ -35$^o$ 44Õ 20.Ó (B1950); V$_{lsr}$=-6.0.
}
   \label{fig2}
\end{center}
\end{figure}



\begin{thebibliography}{}

 \bibitem[Greason86]{1986MST} Greason, M.~R.\ 1986, \textit{M.S.~Thesis}, U. of Va.  

 \bibitem[Jagod et al.(1992)]{1992JMoSp...153..666B} Jagod, 
M., Roesslein, M., Gabrys, C.~M., \& Oka, T.\ 1992, \textit{Journal of Molecular 
Spectroscopy}, 55, 153, 666 
 
 \bibitem[Ross91]{1991ApJ...382L..51R} Roesslein, M., Jagod, 
M., Gabrys, C.~M., \& Oka, T.\ 1991, \textit{ApJ} (Letters), 382, L51 

\bibitem[Roueff07)]{2007A&A...464..245R} Roueff, E., Parise, B., \& Herbst, E.\ 2007, \textit{AAp}, 464, 245 


\bibitem[Wootten87]{1987IAUS..120..311W} {Wootten, A.} 1987, in: {Vardya}, M.~S. and {Tarafdar}, S.~P. (eds.), 
 \textit{IAU Symposium 120, Astrochemistry} (Dordrecht, D. Reidel Publishing Co.), p. 311

\bibitem[WoottenLorSn82]{1982ApJ...255..160W} Wootten, A., Loren, 
R.~B., \& Snell, R.~L.\ 1982, \textit{ApJ}, 255, 160  

\end{thebibliography}
\end{document}